# Multibondic cluster algorithm

Wolfhard Janke and Stefan Kappler[a]*

[a]Institut für Physik, Johannes Gutenberg-Universität Mainz,
Staudinger Weg 7, D-55099 Mainz, Germany

Inspired by the multicanonical approach to simulations of first-order phase transitions we propose for $q$-state Potts models a combination of cluster updates with reweighting of the bond configurations in the Fortuin-Kastelein-Swendsen-Wang representation of this model. Numerical tests for the two-dimensional models with $q = 7, 10$ and $20$ show that the autocorrelation times of this algorithm grow with the system size $V$ as $\tau \propto V^\alpha$, where the exponent takes the optimal random walk value of $\alpha \approx 1$.

## 1. INTRODUCTION

Monte Carlo simulations of first-order phase transitions [1] in the canonical ensemble are severely hampered by extremely large autocorrelation times $\tau \propto \exp(2\sigma L^{D-1})$ where $\sigma$ is the (reduced) interface tension between the coexisting phases and $L^{D-1}$ is the cross-section of the system. To overcome this problem Berg and Neuhaus [2] have recently introduced multicanonical simulations which are based on reweighting ideas and can, in principle, be combined with any legitimate update algorithm. Using *local* update algorithms (Metropolis or heat-bath) it has been demonstrated in several applications [3] that the growth of autocorrelation times with system size is reduced to a power-law, $\tau \propto V^\alpha$ with $\alpha \geq 1$. For the two-dimensional ($2D$) $q$-state Potts model values of $\alpha \approx 1.3$ have been reported for $q = 7$ [4] and $q = 10$ [5].

Since by construction the multicanonical energy distribution is constant over the interesting energy range, a random walk argument would imply an exponent $\alpha = 1$ for an optimally designed update algorithm. Here we present for Potts models a variant of the multicanonical approach based on *non-local* cluster updates which turns out to be optimal in this sense [6]. Basically the idea is to treat the cluster flips in the first place and to reweight the bond degrees of freedom instead of the energy.

## 2. ALGORITHM

The basis of cluster algorithms [7] is the equivalence $Z_{\text{Potts}} = Z_{\text{FK}} = Z_{\text{RC}}$ [8] of the spin representation of the Potts model ($\sigma_i = 1, \ldots, q$)

$$Z_{\text{Potts}} = \sum_{\{\sigma_i\}} e^{-\beta E} \; ; \; E = -\sum_{\langle ij \rangle} \delta_{\sigma_i \sigma_j} \qquad (1)$$

with the Fortuin-Kastelein (FK)

$$Z_{\text{FK}} = \sum_{\{\sigma_i\}} \sum_{\{b_{ij}\}} \prod_{\langle ij \rangle} \left[ p \, \delta_{\sigma_i \sigma_j} \delta_{b_{ij},1} + \delta_{b_{ij},0} \right] \qquad (2)$$

and random cluster (RC) representations

$$Z_{\text{RC}} = \sum_{\{b_{ij}\}} p^B q^{N_c(\{b_{ij}\})} \; ; \; B = \sum_{\langle ij \rangle} b_{ij}, \qquad (3)$$

where $p = \exp(\beta) - 1$. Here $b_{ij} = 0$ or $1$ are bond variables and $N_c(\{b_{ij}\})$ denotes the number of clusters. A cluster is a set of sites which are connected by active bonds $b_{ij} = 1$. By differentiating $\ln Z$ with respect to $\beta$ it is easy to see that the average of the energy $E$ can be expressed in terms of the average of the number of active bonds $B$,

$$\frac{\partial \ln Z}{\partial \beta} = -\langle E \rangle = \frac{p+1}{p} \langle B \rangle. \qquad (4)$$

This suggests that the bond histogram $P_b(B)$ should develop for $\beta = \beta_t \pm \delta\beta$ a pronounced peak around $B_{o,d} = -\frac{p}{p+1} E_{o,d}$, and in the vicinity of

*W.J. would like to thank the DFG for a Heisenberg fellowship and S.K. gratefully acknowledges a fellowship by the Graduiertenkolleg "Physik und Chemie supramolekularer Systeme". Work supported by computer grants HLRZ hkf001 and NVV bvpf03.

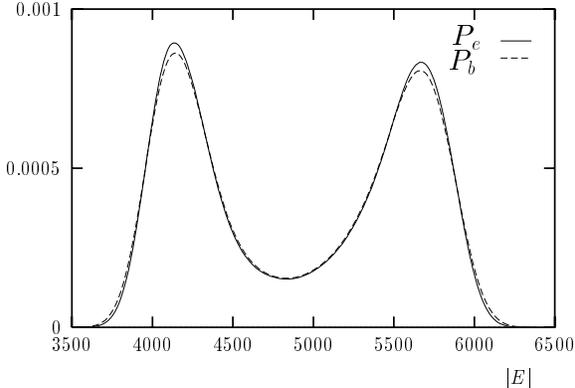

Figure 1. Canonical energy and bond histograms for $q=7$, $L=60$, and $\beta=1.292283$. The bond histogram is plotted vs $[(p+1)/p]B$

the transition point $\beta_t$ a double-peak structure similar to the energy histogram $P_e(E)$. In fact, as is illustrated in Fig. 1 for $q=7$ and $L=60$, a plot of $P_b$ versus $\frac{p+1}{p}B$ is hardly distinguishable from $P_e$ versus $|E|$. For other values of $q$ and $L$ the comparison looks very similar.

In terms of $P_b$ the slowing down of canonical simulations is thus caused by the strongly suppressed configurations between the two peaks, analogous to the well-known argument for $P_e$. To enhance these probabilities we therefore introduce in analogy to multicanonical simulations a "multibondic" partition function

$$Z_{\text{mubo}} = \sum_{\{\sigma_i\}}\sum_{\{b_{ij}\}}\prod_{\langle ij\rangle}\left[p\,\delta_{\sigma_i\sigma_j}\delta_{b_{ij},1}+\delta_{b_{ij},0}\right]w_b(B), \quad (5)$$

where $w_b(B) = P_b^{-1}(B)$ between the two peaks and $w_b(B)=1$ otherwise. Canonical expectation values can always be recovered exactly by applying the inverse of the reweighting factor $w_b(B)$. According to (5) we update the $b_{ij}$ as follows. If $\sigma_i \neq \sigma_j$ then the bond $b_{ij}$ is never active and we always set $b_{ij}=0$. If $\sigma_i=\sigma_j$ then we define $B'=B-b_{ij}$ and choose new values $b_{ij}=0$ or 1 with relative probability $w_b(B') : p\,w_b(B'+1)$. Since for $w_b(B)=1$ we just recover the original Swendsen-Wang algorithm we can now proceed in the usual way by identifying clusters of connected spins and choosing new random values $1\ldots q$ independently for each cluster.

## 3. RESULTS

To evaluate the performance of the multibondic (mubo) cluster algorithm we performed simulations of the $2D$ Potts model (1) with $q=7,10$ and 20 on different lattice sizes for temperatures where the peaks of $P_e(E)$ have approximately the same height [6]. For comparison we ran standard multicanonical (muca) simulations using the heat-bath update algorithm with the same parameters. In each run we recorded 100 000 measurements of $E$ and $B$ in a time-series file. Between the measurements we performed several lattice sweeps to ensure that the autocorrelation times in units of measurements and thus the effective statistics of practically uncorrelated data was roughly the same in all simulations.

For direct comparison with previous work we define flipping times $4\tau_E^{\text{flip}}$ by counting the number of update sweeps that are needed to travel from $E<E_{\min}$ to $E>E_{\max}$ and back. The cuts $E_{\min,\max}$ are chosen as the peak locations $E_{o,d}(L)$ of $P_e(E)$. In our simulations we have tested if $E$ has passed the cuts after each sweep. Our results for $\tau_E^{\text{flip}}$ obtained in multicanonical and multibondic simulations for $q=7$ and 10 are shown in the log-log plots of Fig. 2. Let us first concentrate on the results for $q=7$ where we show for comparison also the data from previous multicanonical simulations [4] and from Rummukainen's hybrid-like two-step algorithm which combines microcanonical cluster updates with a multicanonical demon refresh [9]. Both cluster update versions show qualitatively the same behavior and, for $L>20$, perform much better than the standard multicanonical algorithm. From least-square fits to $\tau_E^{\text{flip}}=aV^\alpha$ we estimate $\alpha\approx 1.3$ for multicanonical heat-bath and $\alpha\approx 1$ for multibondic cluster simulations; see Table 1. Unfortunately, for $q=10$ and 20 the situation is less favorable for the multibondic algorithm. While we still find an exponent of $\alpha \approx 1$, the prefactor $a$ turns out to be so large that we can take advantage of this asymptotic improvement only for very large lattice sizes. As can be seen in Fig. 2, for $q = 10$ the cross-over happens around $L = 50$. Extrapolating to $L = 100$ we estimate that the multibondic algorithm would perform for this lattice size about 1.5 times faster than the standard multicanonical heat-bath. For $q = 20$ the same comparison clearly favors the standard algorithm for all reasonable lattice sizes – and we certainly cannot recommend the new algorithm for large $q$.

Table 1
Results for the dynamical exponent $\alpha$ in multicanonical and multibondic simulations.

|      | $q=7$   | $q=10$  | $q=20$  |
|------|---------|---------|---------|
| muca | 1.27(2) | 1.32(2) | 1.26(1) |
| mubo | 0.92(2) | 1.05(1) | 1.09(1) |

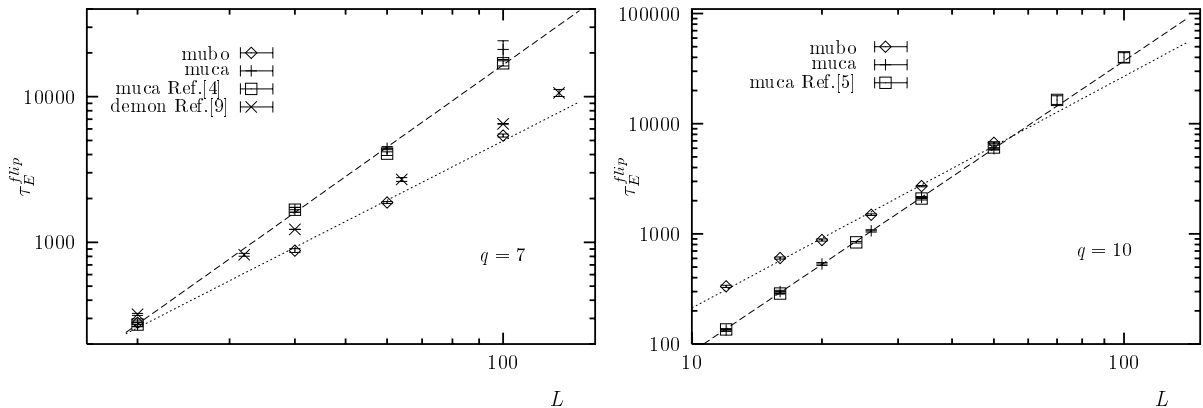

Figure 2. Log-log plots of autocorrelation times $\tau_E^{\text{flip}}$ of the energy vs lattice size for $q = 7$ and 10.

## 4. CONCLUSIONS

The multibondic cluster algorithm provides a combination of cluster update techniques with reweighting in the random bond representation. It is technically not more involved than the multicanonical approach and one lattice sweep takes about the same CPU time. This new algorithm is optimal in the sense that the exponent $\alpha$ is consistent with the optimal random walk value of $\alpha = 1$ and it clearly outperforms the multicanonical heat-bath algorithm. Compared to Rummukainen's intricate algorithm $\tau_E^{\text{flip}}$ is smaller by a factor of $\approx 1.5$. For larger values of $q$, however, the prefactor $a$ turns out to be relatively large, rendering the new algorithm for reasonable lattice sizes more efficient than multicanonical simulations only for $q < q_0$ with $q_0$ somewhat above 10.